**Signature of high $T_c$ around 25K in higher quality heavily boron-doped diamond**

Hiroyuki Okazaki,[1,+] Takanori Wakita,[2] Takayuki Muro,[3] Tetsuya Nakamura,[3] Yuji Muraoka,[2] Takayoshi Yokoya,[2] Shin-ichiro Kurihara,[4] Hiroshi Kawarada,[4] Tamio Oguchi[5] and Yoshihiko Takano[1]

[1]National Institute for Materials Science, Tsukuba, Ibaraki 305-0047, Japan

[2]The Graduate School of Natural Science and Technology, Okayama University, Okayama 700-8530, Japan

[3]Japan Synchrotron Radiation Research Institute (JASRI)/SPring-8, Sayo, Hyogo 679-5198, Japan

[4]School of Science and Engineering, Waseda University, Shinjuku, Tokyo 169-8555, Japan

[5]Institute of Scientific and Industrial Research, Osaka University, Mihogaoka, Ibaraki, Osaka 567-0047, Japan

[+]Present Address: Advanced Institute for Materials Research (AIMR), Tohoku University, Sendai, Miyagi, 980-8577 Japan

**Diamond has outstanding physical properties: the hardest known material, a wide band gap, the highest recorded value for thermal conductivity, and a very high Debye temperature. In 2004, Ekimov *et al*. discovered that heavily boron-doped (B-doped) diamond becomes a superconductor around 4 K.[1] Our group successfully controlled the boron concentration and synthesized homoepitaxially grown superconducting diamond films by a chemical vapor deposition (CVD) method.[2] By CVD method, we found that superconductivity appears when the boron concentration exceeds a critical value of $3.0 \times 10^{20}$ cm$^{-3}$ corresponding to a metal–insulator transition[3] and its $T_c^{zero}$ increases up to 7.4 K with increasing boron concentration.[4-6] We additionally elucidated that the holes formed at the valence band are responsible for the metallic states leading to superconductivity.[7] The calculations predicted that the hole doping into the valence band induces strong attractive interaction and a rapid increase in $T_c$ with increasing**

boron concentration.[8,9] According to the calculations, if substitutional doped boron could be arranged periodically or the degree of disorder is reduced, a $T_c$ of approximately 100 K could be achieved via minimal percent doping. In this work, we have successfully observed zero resistivity above 10 K and an onset of resistivity reduction at 25.2 K in heavily B-doped diamond film. However, the effective carrier concentration is similar to that of superconducting diamond with a lower $T_c$. We found that the carrier has a longer mean free path and lifetime than previously reported,[7,10] indicating that this highest $T_c$ diamond has better crystallinity compared to that of other superconducting diamond films. In addition, the susceptibility shows a small transition above 20 K in the high quality diamond, suggesting a signature of superconductivity above 20 K. These results strongly suggest that heavier carrier doped defect-free crystalline diamond could give rise to high $T_c$ diamond.

To realize high $T_c$ in heavily B-doped diamond we synthesized homoepitaxially grown heavily B-doped high quality diamond films. The thickness is set to 280 nm since a cross-section TEM study reported that nearly perfect diamond with substitutional boron can be deposited until a limit of 500 nm on a (111) diamond substrate.[11] Figure 1(a) shows the temperature dependent resistivity with a clear superconducting transition around 10 K. The zero resistivity is observed at 10.2 K, which is the first report that $T_c^{zero}$ of superconducting diamond exceeds 10 K. In addition, the diamond shows a sharper superconducting transition than previously reported,[4-6] indicating high quality superconducting diamond. In the temperature dependence of magnetic susceptibility applying 10 Oe, a diamagnetic signal is clearly observed at 10.2 K same as our observed $T_c^{zero}$, as shown in Fig. 1(d).

We additionally measured the electronic structure of the heavily B-doped diamond film by soft x-ray angle-resolved photoemission spectroscopy (ARPES), as shown in Fig. 2(a). The ARPES intensities are plotted with respect to the binding energy $E_B$ and momentum $k$. We could measure ARPES spectra including Γ point in Brillouin zone (BZ) using incident photon energy of 835 eV. In Fig. 2(b), the same intensity map is compared with a rigid band shift model of the calculated valence band dispersions for pure diamond and is in good agreement with the calculations. The experimental bandwidth of 23eV is larger than that of calculated value (21eV). Thus the calculated diamond dispersions are energy-enlarged by 10% because of many-body effects on electron-removal energies probed in the photoemission.[12] This is consistent with the previous experimental results of the superconducting (111) diamond films.[6] The band dispersions near $E_F$ are plotted in Fig. 2 (c). Higher intensity bands show clear dispersions toward $E_F$ and Γ point ($k = 0$). These bands clearly cross at Fermi level ($E_F$), as evident from the sudden reduction of intensity at $E_F$ due to the Fermi-Dirac distribution function, indicating the formation of hole pockets in the top of the diamond-like valence band around Γ point. This result is same as previously reported.[7,10] In Fig. 2(d), we compared the experimental intensity map with band structure calculations. The calculated band dispersions were shifted in order to match the Fermi momenta ($k_F$). The derivation of $k_F$ is mentioned below. We found that a location of $E_F$ at 0.5 eV below the top of the valence band. The experimental band dispersions and band shift are similar to that of a previous ARPES intensity map which measured the diamond film with $T_c$ of 6.9 K.[10] From the density of states and the band shift, we estimated the carrier concentration: $n = 3.1 \times 10^{21}$ cm$^{-3}$ (= 1.8 %).

In Fig. 3(a), the ARPES intensities near $E_F$ for several incident photon energies (785-875 eV) were mapped out in the ΓLUX plane using a free-electron final-state model,[7] together with the Fermi momenta and the calculated FS based on rigidly shifted diamond valence band. The higher intensity distribution indicates a Fermi surface (FS). We can confirm band crossing positions corresponding to $k_F$ from the momentum distribution curves (MDCs) at $E_F$. To compare the experimental FS with the calculations, we analyzed the MDCs using Lorentzian functions, as shown in Fig. 3(b). Experimental $k_F$'s were plotted over BZ, together with calculated FS sheets based on the band shift of 0.5 eV. Around Γ point, $k_F$'s are found to coincide with calculated FS sheets, indicating that the carrier concentration resides in $n = 3.1 \times 10^{21}$ cm$^{-3}$. The carrier concentration is close to that of the superconducting diamond film with lower $T_c$ of 6.9 K.[10] However, FS of the highest $T_c$ diamond are clearer than that of previous ARPES study.[6] A clearer FS is expected when the superconducting diamond has a less disordered crystallinity compared to those previously reported. We estimated the mean free path $l$ and the carrier lifetime $\tau$ from the full-width half maximum of the Lorentzian for MDC to be 11.6 Å and 3.8 fs, respectively. Here a comparison with other superconducting diamond samples is important. Table 1 shows the comparison of the present results with previous ARPES results.[7,10] *Previous I* shows heavy carrier than *Previous II* in spite of having a similar $T_c$. Since previous reports shows an estimated error of the carrier concentration, these diamond samples probably have almost the same carrier density. In contrast, present results show little error because of the clearer FS. From the table, we found that $l$ and $\tau$ are longer than those of previous results, which presents good crystallinity of the highest $T_c$ diamond.

The results indicate that higher $T_c$ originates from good crystallinity as the calculations predicted.[8,9]

Moreover, we successfully observed a signature of high $T_c$ in this high quality diamond around 25 K. Figure 4(a) shows the temperature dependence of resistivity under a magnetic field. We observed the separation between the resistivities above 20 K and estimated the onset of transition to be at a value of 25.2 K, which is the highest transition temperature reported so far.[2,5] The onset gradually is shifted to lower temperatures with increasing magnetic field. In addition, we observed a small transition around 25 K close to the onset in the resistivity, as shown in Fig. 4(b), suggesting that a very small fraction shows the superconductivity. If the onset is $T_c^{on}$, the upper critical field ($H_{c2}$) is determined to be 30.1 from the linear extrapolation of the field dependent $T_c^{on}$, as shown in Fig. 4(c). Assuming that the superconductivity in heavily B-doped diamond is the dirty limit, the upper critical field $H_{c2}$ is estimated to be 21.0 T using the equation $H = H_{c2}^{WHH} (1 - (T/T_c)^{1.5})$. We also estimated the irreversible field ($H_{irr}$) to be at a value of 13.0 T. To the best of our knowledge these values are also the highest $H_{c2}$ and $H_{irr}$ reported so far.[5] The signature of higher $T_c$ may be attributed only to better crystallinity or a combination of better crystallinity and partially heavier carrier density. On the other hand, the high temperature superconductivity may also arise at the interface by the correlation between the carrier in B-doped diamond side and the phonon of the pure diamond substrate side with higher frequency than that of heavily B-doped diamond. The signature above 20 K should be already reported if the small high $T_c$ region arises as a consequence of only partially heavier carrier density or only the interface effect. Thus, we expect that the good crystallinity is important for the small high $T_c$ region.

The present results demonstrate that high quality heavily B-doped diamond film shows superconductivity with the highest $T_c^{zero}$ reported so far. Since this diamond film shows clearer FS and longer mean free path and carrier lifetime but has a similar carrier density to previous diamond film with $T_c$ of 6.9 K, less crystal disorder induces higher $T_c$. Additionally we observed the signature of a transition around 25 K in both the resistivity and the susceptibility. These results suggest that high $T_c$ diamond could be produced through a perfect crystalline diamond with higher carrier density.

**Methods**

Homoepitaxially grown heavily B-doped single-crystal (111) diamond film was synthesized on a diamond (111) substrate using a MPCVD method as described in ref. 2-7. Resistivity measurements were carried out using a standard four-probe method in a physical property measurement system (PPMS, Quantum Design). We also confirmed $T_c$ of the film from the magnetic susceptibility by a superconducting quantum interface device (SQUID, Quantum Design).

High-resolution soft x-ray angle resolved photoemission spectroscopy (ARPES) measurements were performed at BL25SU, SPring-8, on a spectrometer built using a Scienta SES200 electron analyzer. ARPES measurements were performed at 9 K and below $2 \times 10^{-8}$ Pa. The incident photon energies were set to be 785-875 eV in order to observe the Fermi surface of superconducting diamond. The total energy resolution was set to be 250 meV. $E_F$ of the sample was referenced to that of a Au film which was measured frequently during the experiments. The ARPES measurements were done after annealing at 700°C under the ultrahigh vacuum to reduce oxygen-related contaminations on the surface.

Band calculations were performed using the first-principles full-potential linearized augmented plane-wave method within local-density approximation. Methodological details follow those used in previous calculations.[7]

**Acknowledgements**

This work was partly supported by Advanced Low Carbon Technology Research and Development Program (ALCA) in Japan Science and Technology Agency (JST). The SXPES measurements at SPring-8 were performed under proposal number 2013A1324.

**Figure captions**

Fig. 1. (a) Temperature dependence of resistivity for heavily B-doped diamond film. (b) Temperature dependence of magnetic susceptibility for both zero-field-cooling (ZFC) and field cooling (FC) processes at a magnetic field of 10Oe.

Fig. 2. (a) ARPES intensity map from heavily B-doped diamond film using incident energy of 835 eV. (b) Comparison of the ARPES intensity map with calculated band dispersions of diamond energy-expanded 10 %. (c) ARPES intensity map near $E_F$. (d) Comparison of the same intensity map of (c) with calculated band dispersions. The calculated band dispersions are shifted by 0.5 eV in order to fit the estimated Fermi momenta from MDC analysis.

Fig. 3. (a) A mapping of ARPES intensities at $E_F$ for incident energy between 785 and 875 eV, together with the experimentally Fermi momenta (solid circles) and the calculated FS sheets (solid lines). (b) MDC spectrum at $E_F$ with four Lorentzian functions. Red circles are the experimental data, blue line is the fitting results, and green lines are the Lorentzian functions used for the fitting. The peak positions correspond to $k_F$.

Fig. 4. (a) Comparison of Temperature dependence of resistivity between 0 T and 7 T. The broken line corresponds to $T_c^{on}$. (b) Extended view of smoothed susceptibility around 25 K of Fig. 1(b). (c) The magnetic field dependence of $T_c^{on}$ and $T_c^{zero}$.

Table 1. Mean free path $l$, lifetime $\tau$, $T_c$, and carrier concentration $n$ estimated from the present and previous ARPES results.[7,10] For ref. 10, we calculated $l$ and $\tau$ from the previous results.

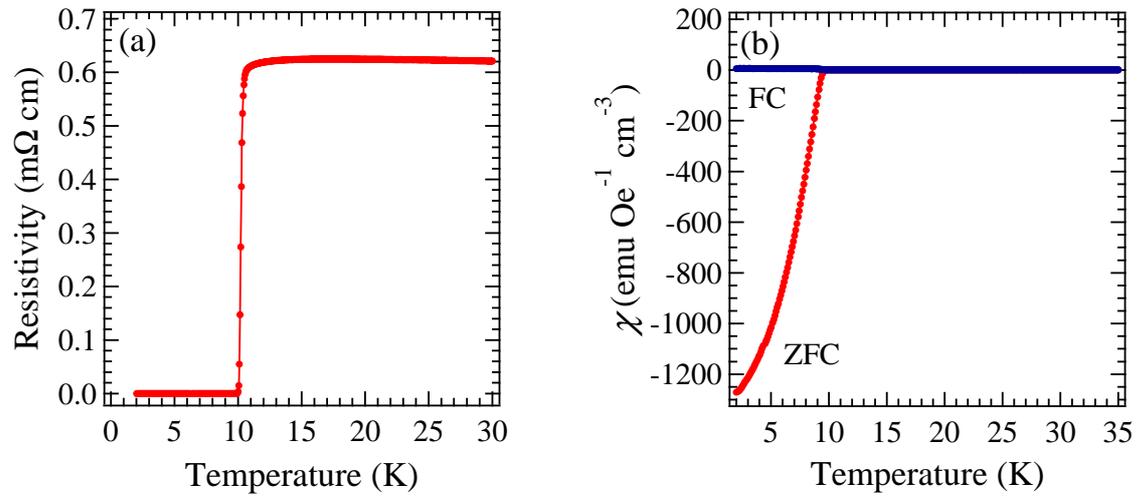

Fig.1  H. Okazaki

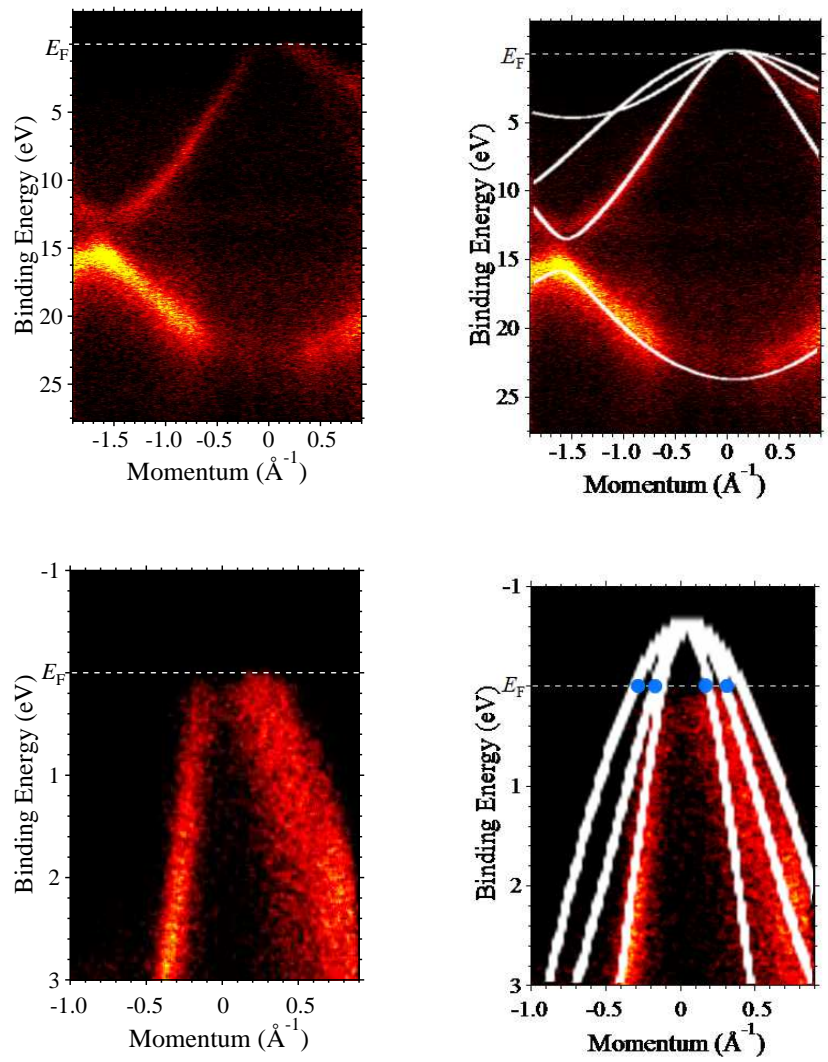

Fig. 2   H. Okazaki

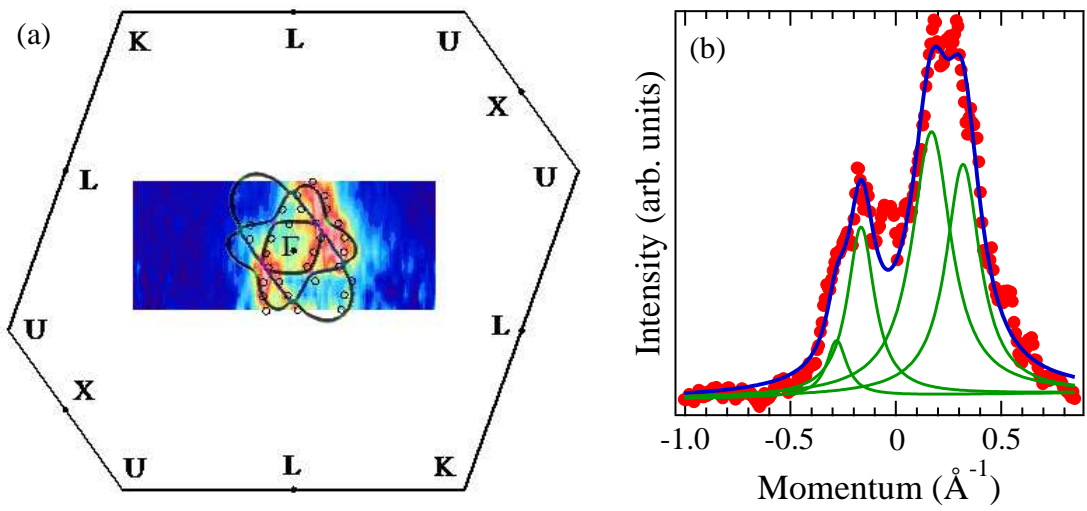

Fig. 3 H. Okazaki

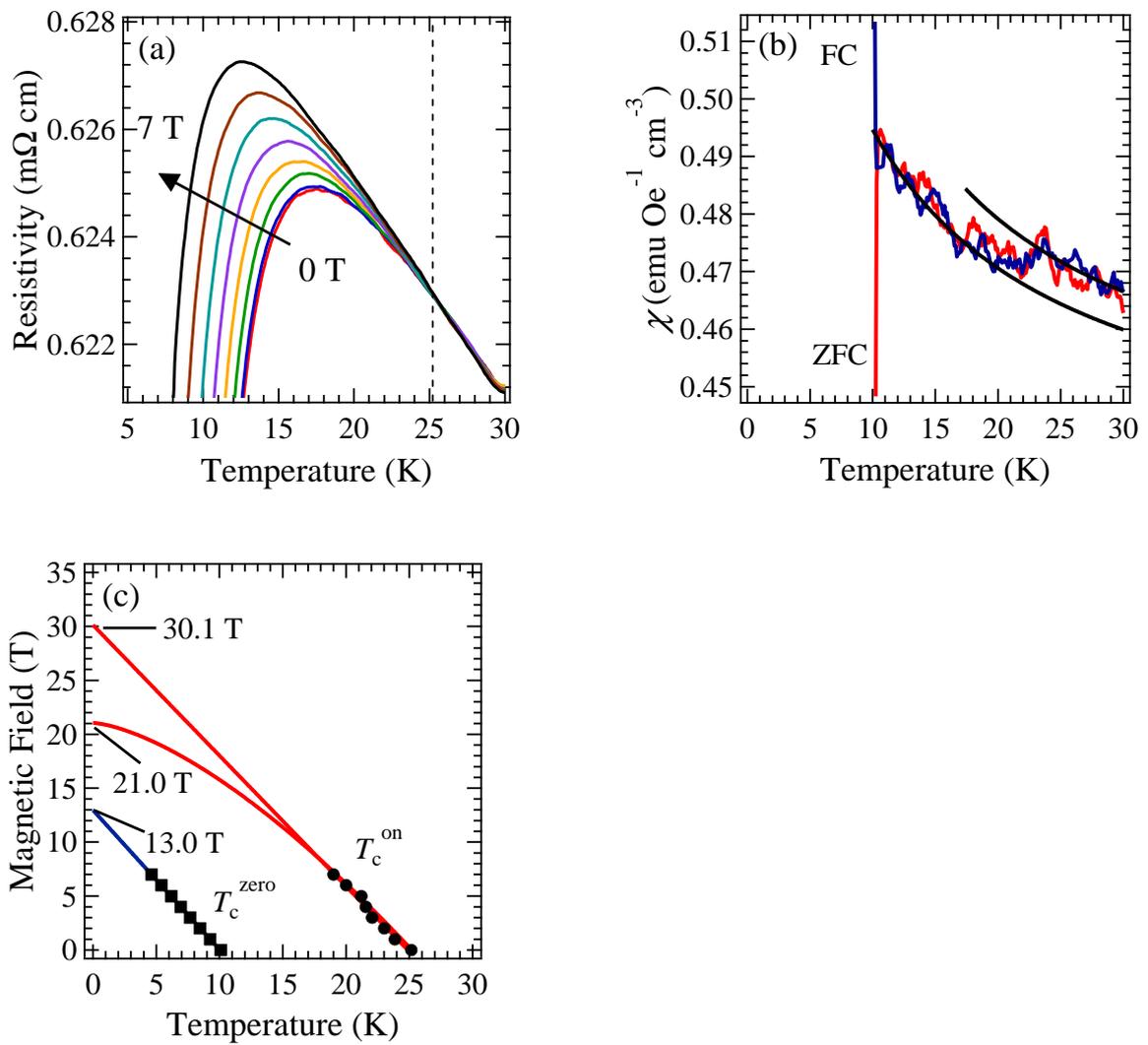

Fig. 4. H. Okazaki

|  | $l$ (Å) | $\tau$ (fs) | $T_c$ (K) | $n$ (cm$^{-3}$) | ref |
|---|---|---|---|---|---|
| Present | 11.6 | 3.8 | 10.2 | $3.1 \times 10^{21}$ | - |
| Previous I | 7.6 | 3.1 | 6.9 | $2.8 \times 10^{21}$ | 7 |
| Previous II | 5.3 | 2.8 | 7.0 | $1.9 \times 10^{21}$ | 10 |

Table 1   H. Okazaki